\documentclass[aps,showpacs,onecolumn,nofootinbib]{revtex4}
\usepackage{graphicx}
\usepackage{dcolumn}
\usepackage{bm}
\def\lsim{\raisebox{-.4ex}{$\stackrel{<}{\scriptstyle \sim}$\,}}
\def\gsim{\raisebox{-.4ex}{$\stackrel{>}{\scriptstyle \sim}$\,}}

\usepackage{latexsym}
\begin{document}
\input epsf
\newcommand{\A}{{\mathcal{A}}}
\newcommand{\dA}{\delta{\mathcal{A}}}
\newcommand{\oR}{\Omega_r h^2}
\newcommand{\og}{\Omega_{\gamma} h^2}
\newcommand{\om}{\Omega_m h^2}
\newcommand{\ok}{\Omega_k h^2}
\newcommand{\ob}{\Omega_b h^2}
\newcommand{\ode}{\Omega_{de} h^2}
\newcommand{\h}{\mathcal{H}}


\title{Perturbations in electromagnetic dark energy}
\author{Jose Beltr\'an Jim\'enez$^a$, Tomi S. Koivisto$^b$,
Antonio L. Maroto$^a$ and David F. Mota$^c$}
\affiliation{$^a$Departamento de  F\'{\i}sica Te\'orica,
 Universidad Complutense de
  Madrid, 28040 Madrid, Spain.\\
 $^b$ Institute for Theoretical Physics, University of Heidelberg, 69120 Heidelberg,Germany
\\
 $^c$Institute of Theoretical Astrophysics, University of Oslo, 0315 Oslo,
 Norway}

\begin{abstract}
It has been recently proposed that the presence of a temporal
 electromagnetic field on cosmological scales could explain the phase of
accelerated expansion that the universe is currently undergoing.
The field contributes as a cosmological constant and therefore,  the
homogeneous cosmology produced by such a model is exactly the
same as that of $\Lambda$CDM .
However, unlike a cosmological constant term, electromagnetic
fields can acquire perturbations which in principle could
affect CMB anisotropies and structure formation. In this work, we
study the evolution of inhomogeneous scalar
perturbations in this model. We show that provided the initial electromagnetic
fluctuations generated during inflation are small, the model is perfectly compatible with both
CMB and large scale structure observations at the same level of accuracy as $\Lambda$CDM.
\end{abstract}

\pacs{98.80.-k, 95.36.+x, 14.70.Bh,04.62.+v}

\date{\today}

\maketitle

\section{INTRODUCTION}

Since the accelerated expansion of the universe was discovered
more than a decade ago \cite{SN1998}, a big effort has been made
in trying to identify the culprit responsible for it. The simplest
explanation is provided by the introduction of a cosmological
constant term in the Einstein equations. In fact, such a solution
is in excellent agreement with most of the cosmological
observations to date and only requires the introduction of one
extra parameter to describe the energy density associated to the
cosmological constant. Hence, as a phenomenological model to
describe the observable universe, it is very successful. However,
the value of the cosmological constant inferred from the
observations is extremely tiny as compared to the other scale
present in the gravitational action and set by the Newton's
constant, namely the Planck scale. This poses a problem from the
theoretical point of view because a theory containing two
dimensional constants which differ by so many orders of magnitude
does not seem to be very {\it natural}. This problem is usually
referred to as the naturalness problem or the cosmological
constant problem, and it is related to the coincidence problem,
that is, why dark energy and matter have similar density
parameters today when their respective energy densities have
evolved very differently throughout the expansion history of the
universe. These two problems are the main reason why so many
models to explain the current acceleration of the universe have
been proposed. Among those models, one can generally find models
in which the accelerated expansion is caused by either a new field
(dark energy) or a modification of Einstein gravity. The popular
quintessence models \cite{quintessence} with a cosmological scalar
field playing the role of dark energy, and their generalizations
with non-canonical kinetic terms, like the k-essence models
\cite{k-essence} or with a coupling to curvature invariants
\cite{gauss} (see \cite{Copeland} for a review on dark energy
models) they all belong to the first group. In the second group we
have the very popular $f(R)$ theories \cite{f(R)} or the models
with extra-dimensions \cite{DGP}. Although all these models were
motivated by the cosmological constant problem, the actual
situation is that most of them fail in solving it because they
rather shift it to some other parameter of the theory. Besides,
they are usually plagued by either classical or quantum
instabilities or present conflicts with local gravity tests
\cite{durrer}.

Although the scalar fields have become the most popular
candidates, vector fields have also been shown to be compelling
candidates for dark energy \cite{VectorDE,vectorlss} and even
solve the coincidence problem \cite{cosmicvector}. Generalisations
to higher forms have also been considered recently \cite{forms}.
Such models have been overlooked since they generically induce an
anisotropy. This picture has changed in the last few years with
the recent detections of some unexpected features in the CMB
temperature anisotropies: A hemispherical asymmetry has been
reported \cite{c4}; the angular correlation spectrum seems to be
lacking power at the largest scales \cite{c3}; the alignment of
the quadrupole and octupole (the so called Axis of Evil
\cite{c2,c1}) could also be seem as an extra-ordinary and unlikely
result of statistically isotropic perturbations, even without
taking into account that these multipoles happen also to be
aligned to some extent with the dipole and with the equinox. The
significance of the anomalies has been debated extensively in the
literature (see e.g. \cite{f1,f2,f3,f4,s1,s2,s3,s4}) with some
reported effects more significant than others
\cite{frode1,frode2,nic}.

One model that, not only succeeds in addressing the cosmological
constant problem, but also is free from both classical and quantum
instabilities and has the same set of PPN parameters as General
Relativity is the one recently proposed in \cite{EMdarkenergy}. In
that model, the role of the cosmological constant is played by a
temporal component of the electromagnetic field\footnote{We refer
here to the vector $A_\mu$ as the electromagnetic field, instead
of calling $A_\mu$ the vector potential and $F_{\mu\nu}$ the
photon.} or, in other words, by the existence of an absolute
cosmological electric potential \cite{AbsPot}. That way, in
addition of solving the aforementioned problems, we can establish
the true nature of dark energy without resorting to unknown
physics. In such a model, a slightly modified theory of the
electromagnetic interaction is proposed and whose motivation comes
from the quantization of the electromagnetic field in the
covariant formalism. In that formalism, one modifies the
electromagnetic action by introducing a gauge fixing term in the
original action in exchange for either imposing the Lorenz
condition $\partial_\nu A^\nu=0$ at the classical level or
reducing the corresponding Hilbert space with the weak Lorenz
condition $\partial^\nu A_\nu^{(+)}\vert\phi\rangle=0$ at the
quantum level. However, when moving to an expanding universe, the
Lorenz condition cannot be consistently imposed at all times and,
therefore, this approach becomes inappropriate. For this reason,
in the model proposed in \cite{EMdarkenergy}, the electromagnetic
interaction is quantized without imposing the Lorenz condition.
The price to pay is that we need to introduce a new
electromagnetic scalar mode in the theory so that the
electromagnetic interaction would actually contain three physical
degrees of freedom (see \cite{turner,bertolami} for other interesting consequences
in theories with extra electromagnetic polarizations).
This new scalar mode is however decoupled from
the conserved currents and can only be excited by gravitational
fields. Then, it could be generated during an inflationary phase
in the early universe and the interesting feature of the model is
that, once it becomes super-Hubble, it gives rise to a new
contribution to the action which is like that of a cosmological
constant. Surprisingly enough is the fact that the observed value
of the cosmological constant is naturally achieved if inflation
occurred at the electroweak scale.

Even though this model provides the same expansion history for the
universe as a pure cosmological constant, its different nature
makes possible to have a different evolution for the
inhomogeneities so that it becomes necessary to perform an
analysis of the perturbations to confront it to CMB and LSS data.
In addition, as the observational data seem to be consistent with
a cosmological term, it is then interesting that, by studying
fluctuations, one in principle is able to infer the physical
origin of the cosmological term. Previously the possibility of
anisotropic fluctuations about a constant density term have been
contemplated in the frameworks of viscous fluids and
noncommutative geometry \cite{Rodrigues:2007ny,vectorlss}. Here
the properties of such a cosmological term can be consistently
derived from a fully covariant theory. The presence of
fluctuations can be understood since the constancy of the
cosmological term is just a property of a solution for a component
of a vector field. Uncovering the evolution of the fluctuations is
the aim of the present work, which is organized in the following
way. In Section I we give a short description of the model. In
Section II we write down the background equations and solve them.
Besides, we show how the background evolution gives rise to the
same expansion history as a cosmological constant. Section III is
devoted to the perturbations for the electromagnetic dark energy
model. In particular, we obtain the corresponding equations and
perturbed expressions in the Newtonian gauge and solve them in
some interesting cases. The analogous expressions for the
synchronous gauge are given in an appendix. Finally, in section IV
we use a modified version of the publicly available numerical code
CAMB \cite{Lewis:1999bs} to compute the evolution of the linear
perturbations in the model. The results are summarized and
discussed in the final Section \ref{disc}.

\section{Description of the model}
The proposed action describing the electromagnetic interaction
including the gauge fixing term and current coupling is given by:
\begin{eqnarray}
S=\int d^4x
\sqrt{g}\left[-\frac{1}{4}F_{\mu\nu}F^{\mu\nu}+\frac{\xi}{2}
(\nabla_\mu A^\mu)^2+ A_\mu J^\mu\right]\label{action}
\end{eqnarray}
with $F_{\mu\nu}=\partial_\mu A_\nu-\partial_\nu A_\mu$. Here, the
gauge fixing term is considered as a fundamental piece of the
action on equal footing to the Maxwell term, i.e. electromagnetism
is considered as a gauge non-invariant theory which contains three
physical degrees of freedom, namely the two usual transverse
photons plus an extra scalar mode (despite the gauge
non-invariance, the theory is perfectly consistent as shown in
\cite{EMdarkenergy}). In fact, the gauge fixing term can be seen
as the kinetic term of this new mode and so the parameter $\xi$
can be fixed to $1/3$ to have canonically normalized fields. While
the general $U(1)$ transformations are no longer an exact symmetry
of the theory, it still preserves a residual gauge symmetry given
by $A_\mu\rightarrow A_\mu+\partial_\mu\theta$ provided
$\Box\theta=0$.

The modified Maxwell equations deduced from (\ref{action}) read:
\begin{eqnarray}
\nabla_\nu F^{\mu\nu}+\xi\nabla^\mu(\nabla_\nu A^\nu)=J^\mu
\label{EMeq}
\end{eqnarray}
If we now take the 4-divergence of these equations we obtain:
\begin{eqnarray} \label{box}
\Box(\nabla_\nu A^\nu)=0\label{minimal}
\end{eqnarray}
where we have used the fact that the electromagnetic current is
covariantly conserved. The latter equation shows that $\nabla_\nu
A^\nu$ behaves as a massless scalar field which is decoupled from the
conserved electromagnetic currents, but it is non-conformally
coupled to gravity so that it can be excited by gravitational
fields. Moreover, due to the well-known fact that a massless scalar field
gets frozen on super-Hubble scales for a FLRW universe, the four
divergence of the electromagnetic field will be constant on scales
larger than the Hubble radius, giving rise to a cosmological
constant-like term in the action. On the other hand, for small
scales we have that $\nabla_\nu A^\nu\sim a^{-1}$ so that its
contribution to the Maxwell equations becomes negligible.
Therefore, as long as the four divergence of the electromagnetic
field is super-Hubble, it can play the role of a cosmological
constant and, once it enters the horizon, it is rapidly suppressed
and we recover the usual Maxwell equations with the Lorenz
condition.

It is interesting to note that the divergence of the
electromagnetic field can be seen, at the equations of motion
level, as a conserved current acting as a source of the usual
Maxwell field. To see this, we can write
$-\xi\nabla^\mu(\nabla_\nu A^\nu)\equiv J_{\nabla\cdot A}^\mu$
which, according to (\ref{minimal}), satisfies the conservation
equation $\nabla_\mu J_{\nabla\cdot A}^\mu=0$ and we can express
(\ref{EMeq}) as:
\begin{eqnarray}
\nabla_\nu F^{\mu\nu}=J^\mu_T
\end{eqnarray}
with $J^\mu_T=J^\mu+J^\mu_{\nabla\cdot A}$ and $\nabla_\mu
J^\mu_T=0$. Physically, it means that, while the new scalar mode
can only be excited by means of gravitational fields, once it is
produced it will generally be considered as a source of electromagnetic
fields.

The energy-momentum tensor corresponding to the modified action
(\ref{action}) is the sum of two pieces:
\begin{eqnarray}
 T_{\mu\nu}=T_{\mu\nu}^M+T_{\mu\nu}^\xi
\end{eqnarray}
with
\begin{equation}
T_{\mu\nu}^M=\frac{1}{4}g_{\mu\nu}F_{\alpha\beta}F^{\alpha\beta}-F_{\mu\alpha}F_\nu^{\;\;\alpha}
\end{equation}
the standard energy-momentum tensor corresponding to Maxwell
theory and
\begin{equation}
T_{\mu\nu}^\xi=\frac{\xi}{2}\left[g_{\mu\nu}\left[\left(\nabla_\alpha
A^\alpha\right)^2 +2A^\alpha\nabla_\alpha\left(\nabla_\beta
A^\beta\right)\right] -4A_{(\mu}\nabla_{\nu)}\left(\nabla_\alpha
A^\alpha\right)\right]
\end{equation}
the energy-momentum tensor corresponding to the gauge fixing term.
Notice that, for a residual gauge mode of the form
$A_\mu=\partial_\mu\theta$ with $\Box\theta=0$, the
energy-momentum tensor vanishes identically.

\section{Background equations}

We shall assume a homogeneous and isotropic background described
by the FLRW metric which, in conformal time, is given by:
\begin{equation}
ds^2=a(\tau)^2\left[d\tau^2-\delta_{ij}dx^idx^j\right].
\end{equation}
In order to preserve the homogeneity we need the electromagnetic
field to depend only on time whereas the large scale isotropy
requires the absence of spatial components so that the background
electromagnetic field is merely given by a temporal component
which, in conformal time coordinates, we shall call $\A_0(\tau)$
\footnote{We shall denote by calligraphic letters the components
of the vector field in conformal time coordinates}. Moreover, this
homogeneous temporal component evolves according to the zero
component of the equations (\ref{EMeq}), that yield:
\begin{equation}
\A_0''+2\left(\h'-2\h^2\right)\A_0=0\label{Backeq}
\end{equation}
with $'\equiv\frac{d}{d\tau}$ and $\h\equiv\frac{a'}{a}$ the
Hubble expansion rate in conformal time. Notice that, since
$F_{\mu\nu}=0$ for the assumed background configuration, the
latter equation simply expresses that $d(\nabla_\nu
A^\nu)/d\tau=0$ which, in turn implies that the four divergence of
the electromagnetic field is constant irrespectively of the
evolution of the scale factor.

If we use now the fact that $F_{\mu\nu}=0$ together with
$\nabla_\nu A^\nu=const$ in the expression for the energy momentum
tensor, we obtain that:
\begin{equation}
T_{\mu\nu}=\frac{\xi}{2}\left(\nabla_\mu A^\mu\right)^2 g_{\mu\nu}
\end{equation}
which is exactly the same as that of a cosmological constant of
value $\Lambda=\frac{\xi}{2}\left(\nabla_\nu A^\nu\right)^2$.
Notice that in order to have a positive cosmological constant we
need $\xi>0$. The background evolution for this model thus mimics
that of the standard $\Lambda$CDM model whose cosmological
constant value is determined by the four divergence of the
electromagnetic field. The energy density of the electromagnetic
field in this configuration is:
\begin{equation}
\rho_{\A_0}=\frac{1}{6a^4}\left(\A_0'+2\h\A_0\right)^2\label{Backrho}
\end{equation}
and the equation of state is simply $p_{A_0}=-\rho_{A_{0}}$. For
the subsequent calculations, it will be convenient to introduce a
redefinition of the vector field as $A_0=a^2\A_0$ so that
expressions (\ref{Backeq}) and (\ref{Backrho}) look much simpler:
\begin{eqnarray}
\left(a^{-4}A_0'\right)'=0\\
\rho_{A_0}=\frac{1}{6a^8}(A_0')^2\label{Backrho2}
\end{eqnarray}
and we clearly see that the energy density is constant.

Finally, the fact that the background vector field does not
contain spatial components is justified because on super-Hubble
scales the spatial components grow more slowly than the temporal
one. Let us see this in more detail. For a configuration with
spatial components of the electromagnetic field in addition to the
temporal one, i.e., with $\A_\mu=(\A_0(\tau),\vec \A (\tau))$ and
in a FLRW universe, the modified Maxwell equations read:
\begin{eqnarray}
\A_0''+2(\h'-2\h^2)\A_0&=&0\\
\vec{\A}\;''&=&0
\end{eqnarray}
Thus, if we assume a power law expansion with $\h=p/\tau$, the
solution of these equations is:
\begin{eqnarray}
\A_0&=&C_{01}\tau^{1+p}+C_{02}\tau^p\\
\vec{\A}&=&\vec{C}_1\tau+\vec{C}_2
\end{eqnarray}
and we see that the spatial components decay with respect to the
temporal one in both the radiation ($p=1$) and matter ($p=2$)
dominated eras. In addition, the energy density associated to the
spatial part decays as $a^{-4}$ whereas that associated to the
temporal component is constant, as we have already seen. Hence, it
makes sense to assume the aforementioned isotropic background with
vanishing spatial components given above.

\section{Scalar perturbation equations}


In this section we shall derive the equations for the
perturbations and solve them analytically for some simple cases.
We only consider the case of scalar perturbations because we have
checked that the vector perturbations evolve in the same way as in
standard Maxwell theory and the tensor perturbations remain
unaffected by the presence of the gauge fixing term as well. One
might expect this because the gauge fixing term only affects the
new scalar mode which, being scalar, can contribute to the scalar
perturbations, but not to the pure transverse vector perturbations
(i.e., the usual photons).  We shall do the calculations of this
section in the conformal Newtonian gauge, although we give the
corresponding expressions for the synchronous gauge in the
appendix.

In principle, the effect of the high-electric conductivity of the
universe should be taken into account including the corresponding
electromagnetic current in the r.h.s  of Maxwell equations. The
reason why we can neglect such effects is the following: the
electromagnetic current should satisfy the conservation equation:
\begin{eqnarray}
\nabla_\mu J^\mu=0
\end{eqnarray}
and also the condition of electric neutrality, i.e.
\begin{eqnarray}
u^\mu J_\mu=0
\end{eqnarray}
with $u_\mu$ the four-velocity of the comoving observers. Let us
expand also the current up to first order as:
\begin{eqnarray}
J_\mu=J_\mu^0+\delta J_\mu
\end{eqnarray}
but $J_\mu^0=0$ for the homogeneous and isotropic electrically
neutral background (this is the reason why we did not consider the
current term in the background equations). If we now assume that
the universe remains neutral at first order in the perturbations
we have that $\delta J_0=0$ and, finally, from current
conservation, we get: $\vec \nabla \cdot\vec{\delta J}=0$, i.e.
the perturbed current is transverse. In other words, when
computing scalar perturbations, we can ignore the effect of
electric conductivity by assuming that the electric charge
vanishes at first order as well.

In the considered gauge, the perturbed line element is given by:
\begin{equation}
ds^2=a(\tau)^2\left[(1+2\psi)d\tau^2-(1-2\phi)\delta_{ij}dx^idx^j\right]
\end{equation}
In the absence of any anisotropic stress sources we have
$\phi=\psi$. In order to simplify the expressions we will use
latter, we define the scalar perturbation of the vector field by
introducing a convenient power of the scale factor as we did for
the background field:
\begin{equation}
\dA_\mu=a^{-2}(\dA_0,\nabla\delta\A)\label{defpert}
\end{equation}
As usual, we shall introduce the Fourier components of the
perturbations and solve the corresponding equations for them.
Then, from (\ref{EMeq}) we can obtain the following equations for
the Fourier modes of the vector field perturbations:
\begin{eqnarray}
&&\dA_{0k}''-4\h\dA_{0k}'-3k^2\dA_{0k}=-2k^2\left(2\dA_k'-5\h\dA_k\right)+
\left[\left(\psi_k''-4\h\psi_k'+3\phi_k''-12\h\phi_k'\right)A_0+3(\psi_k'+\phi_k')A_0'\right]\nonumber\\
&&\dA_k''-4\h\dA_k'+\left(4\h^2-2\h'-\frac{1}{3}k^2\right)\dA_k=
2\left(\frac{2}{3}\dA_{0k}'-\h\dA_{0k}\right)-\left[\left(\frac{1}{3}\psi_k'+\phi_k'\right)A_0+\frac{2}{3}\psi_k
A_0'\right]\label{perteqNewt}
\end{eqnarray}
In these equations we can see that the two perturbations of the
vector field are coupled to each other and that the metric
perturbations act as a source of them. Then, even in the case that
the initial perturbations vanish the gravitational potentials will
generate perturbations of the vector field that, eventually, may
also be source of the metric perturbations.

On the other hand, the corresponding perturbed energy-momentum
tensor components are given by:
\begin{eqnarray}
\delta T^0_{\;\,0}&=&\frac{1}{3a^8}\left\{ [-2\psi_k
A_0'+\dA_{0k}'-(3\phi_k'+\psi_k')A_0]A_0'+
k^2\left[(3\dA_k'-6\h\dA_k-3\dA_{0k})A_0+\dA_k A_0'\right]\right\}\nonumber\\
\delta T^i_{\;\,j}&=& \frac{1}{3a^8}\left\{ [-2\psi_k
A_0'+\dA_{0k}'-(3\phi_k'+\psi_k')A_0]A_0'+
k^2\left[(-3\dA_k'+6\h\dA_k+3\dA_{0k})A_0+\dA_k A_0'\right]\right\}\delta^i_{\;\,j}
\nonumber\\
\delta
T^0_{\;\,i}&=&-\frac{ik_i}{3a^8}A_0\left[-(3\phi_k'+\psi_k')A_0-2\psi_k
A_0' +\dA_{0k}'+k^2\dA_k\right]\label{deltaTconf}
\end{eqnarray}
It is interesting to note that this model has vanishing shear,
i.e., $\delta T^i_{\;\,j}=-\delta p_k\;\delta^i_j$. This is due to there
being only one physical scalar mode present. Moreover, from
expressions (\ref{deltaTconf}) one can find the following relation
between the perturbed energy density and pressure of the field:
\begin{equation}
\delta(\rho_k+ p_k)=-\frac{2
A_0k^2}{a^8}\left(\dA_{0k}-\dA_k'+2\h\dA_k\right)\label{masspert}
\end{equation}
This relation is important because for a gauge mode satisfying
$a^{-2}\dA_\mu=\partial_\mu \chi$ (remember here the definition of
the perturbations given in (\ref{defpert})) one has that
$\dA_0=\dA'+2\h\dA$ and, as a consequence, $\delta(\rho_k+p_k)=0$.

\begin{figure}
\begin{center}
{\epsfxsize=15.0 cm\epsfbox{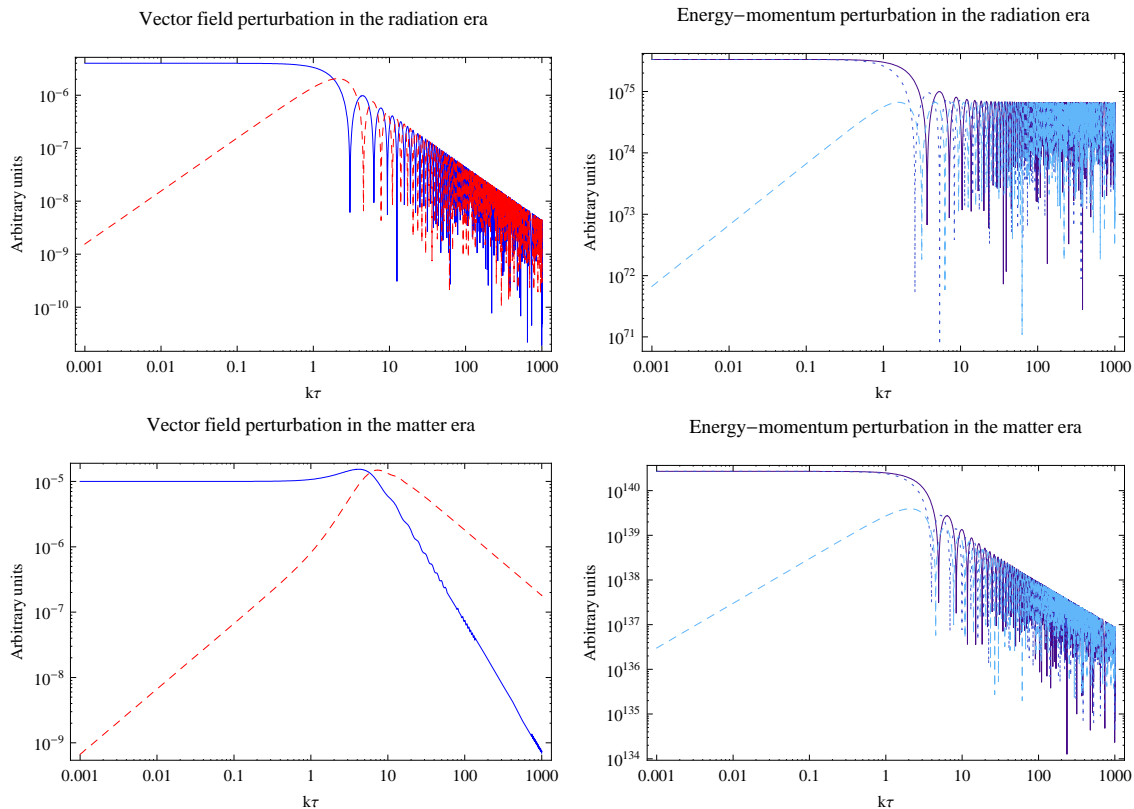}} \caption{ In these
figures we show the evolution of the vector field perturbations in
the Newtonian gauge for radiation (upper panel) and matter (lower
panel) dominated universes. In the left panels we show the
evolution of the vector field perturbations $\dA_0/A_0$
(solid-blue) and $k\dA/A_0$ (dashed-red). The right panels show
the perturbed energy density $\delta T^0_0$ (solid), pressure
$-\frac{1}{3}\delta T^i_i$ (dotted) and momentum
$-\frac{1}{ik^2}k_i\delta T^0_i$ (dashed).
}\label{pertnewt}\end{center}
\end{figure}

In order to obtain some analytical results, in the following we
shall consider that the metric perturbations are generated by some
dominating fluid and study the evolution of the electromagnetic
perturbations in such a scenario. In other words, we shall assume
that the perturbations of the electromagnetic field will not
affect the metric perturbations evolutions. This assumption is
justified as long as the electromagnetic energy density is clearly
subdominant as it happens in most of the universe evolution when
the energy density associated to the electromagnetic field is many
orders of magnitude below that of the dominant component. However,
such a condition will eventually breakdown at low redshift when
dark energy becomes dominant and the results obtained here lack
validity, being necessary to resort to a numerical treatment. In
the early universe when radiation represents the dominant
contribution to the energy density of the universe and neglecting
neutrinos shear (which implies that $\psi=\phi$) the metric
perturbation evolve as \cite{Mukhanov}:
\begin{equation}
\phi_k=\frac{C_{1k}[\omega\tau\cos(\omega\tau)-\sin(\omega\tau)]+
C_{2k}[\omega\tau\sin(\omega\tau)+\cos(\omega\tau)]}{\tau^3}
\end{equation}
with $\omega=k/\sqrt{3}$. On the other hand, in the matter
dominated universe, the gravitational potential becomes constant
in time, i.e., $\phi_k=const$. Then, we can solve the equations
(\ref{perteqNewt}) in the presence of the gravitational
perturbations produced by a radiation or matter fluid and obtain
the evolution of the vector field perturbations in those epochs.
The results are shown in Fig. \ref{pertnewt}. We can see that, on
super-Hubble scales, the perturbation $\dA_0$ evolves in the same
way as the background vector field so that $\dA_0/A_0=const$ as
one would expect. This also implies that the perturbed energy
density is constant on large-scales as we can see in the figure.
Moreover, this feature does not depend on the dominating fluid,
i.e., it happens for both the radiation and matter eras. On small
scales, the perturbed energy density scales with constant
amplitude when the universe is dominated by radiation whereas the
amplitude decays as $1/\tau$ in the matter era. Notice that this
behavior on small scales is a common feature for all the perturbed
components of the energy-momentum tensor, i.e., the energy, the
pressure and the scalar component of the momentum.

We shall obtain the explicit evolution for the electromagnetic
perturbations during the matter-dominated era, when most of the
cosmologically relevant scales reenter the horizon. In that epoch,
the gravitational potential is constant as we said above so that
$\phi_k=\psi_k=\phi_0$ and the Hubble parameter satisfies
$\h'=-\frac{1}{2}\h^2$. With these conditions, we can obtain the
following expression for $\dA$ in terms of $\dA_0$ and $\phi_0$:
\begin{eqnarray}
\dA_k=-\frac{3}{4k^2}\left[\dA_{0k}'''-\frac{11}{2}\h\dA_{0k}''+
\left(\frac{7}{3}k^2-8\h^2\right)\dA_{0k}'-\frac{7}{2}\h
k^2\dA_{0k} -\frac{8}{3}A_0'\phi_0\right]\label{dA0kevol}
\end{eqnarray}
This relation allows to find the following fourth-order
differential equation for $\dA_{0k}$:
\begin{eqnarray}
\dA_{0k}^{\textrm{iv}}-8\h\dA_{0k}'''+2\left(k^2-\frac{49}{4}\h^2\right)\dA_{0k}''-
8\h^2\left(k+\frac{7}{2}\h\right)\dA_{0k}'+k^2\left(k^2+\frac{21}{2}\h^2\right)\dA_{0k}=
4k^2\h A_0'\phi_0\label{forthoreq}
\end{eqnarray}
This equation together with the relation (\ref{dA0kevol})
determine the evolution of the electromagnetic perturbations in a
matter dominated universe. It is convenient to remind here the
assumptions under which such equations remain valid. In order to
obtain those equations we have assumed that the metric
perturbations act as an external source for the electromagnetic
perturbations and that this external source is uniquely determined
by the matter fluid. This means that the contribution of the
electromagnetic field to the perturbed Einstein equations are
negligible with respect to that of the matter component, which is
a good approximation as long as the electromagnetic field energy
density is well below the matter energy density. However, this
condition eventually breakdowns because the electromagnetic field
becomes dominant and it contributes in a non-negligible way to the
Einstein equations so that the metric perturbations becomes
affected by the electromagnetic field perturbations and the full
system of coupled equations must be solved.

In order to solve eq. (\ref{forthoreq}) we shall take advantage of
the residual gauge symmetry of the theory, namely,
$A_\mu\rightarrow A_\mu+\partial_\mu\theta$ with $\Box\theta=0$.
From this symmetry, we know that $\dA_0=a^2\theta'$ will be
solution of (\ref{forthoreq}) so that we can factorize it as
${\mathcal{F}}[{\mathcal{G}}(\dA_{0})]=4k^2\h A_0'\phi_0$, where
$\mathcal{F}$ is a second order differential operator and
$\mathcal{G}$ is the operator determining the evolution of
$a^2\theta'$ which can be deduced from the equation satisfied for
$\theta$ and turns out to be
$\mathcal{G}=\frac{d^2}{d\tau^2}-2\h\frac{d}{d\tau}+k^2$.
Therefore, the equation (\ref{forthoreq}) can be expressed as:
\begin{equation}
\left[\frac{d^2}{d\tau^2}-6\h\frac{d}{d\tau}+\left(k^2+\frac{21}{2}\h^2\right)
\right]\left[\frac{d^2}{d\tau^2}-2\h\frac{d}{d\tau}+k^2\right]\dA_{0k}=4k^2\h
A_0'\phi_0
\end{equation}
Thus, the solution for $\dA_0$ will be determined by the equation
${\mathcal{G}}(\dA_0)=S$ with $S$ the solution of the equation
${\mathcal{F}}(S)=4k^2\h A_0'\phi_0$. Since $\h=\frac{2}{\tau}$ in
the matter era, we can obtain the explicit form of the kernel of
$\mathcal{F}$ (solutions of the homogeneous equation) and is given
by:
\begin{equation}
S_{hom}(\tau)=\tau^6\left(C_{S1}e^{-ik\tau}+C_{S2}e^{ik\tau}\right)
\end{equation}
whereas a particular one can be obtained by:
\begin{equation}
S_{part}=S_1\int\frac{S_2}{\det W_S}4k^2\h A_0'\phi_0d\tau-
S_2\int\frac{S_1}{\det W_S}4k^2\h A_0'\phi_0d\tau
\end{equation}
where $S_1$ and $S_2$ are the two independent solutions given
above and $\det W_S= S_1'S_2-S_2'S_1=2ik\tau^{12}C_{S1}C_{S2}$ is
the determinant of the Wronskian. Following the same procedure, we
obtain the kernel of ${\mathcal{G}}$
\begin{equation}
\dA_0^{hom}(\tau)=C_1\left(k^2\tau^2-3ik\tau-3\right)e^{-ik\tau}
+C_2(k^2\tau^2+3ik\tau-3)e^{ik\tau}\label{homsolu}
\end{equation}
whereas the particular solution is
\begin{equation}
\dA_0^{part}=\dA_{0,C_1}\int\frac{\dA_{0,C_2}}{\det
W_C}S(\tau)d\tau-\dA_{0,C_2}\int\frac{\dA_{0,C_1}}{\det
W_C}S(\tau)d\tau
\end{equation}
with $\dA_{0,C_i}$ $(i=1,2)$ denoting the piece of the homogeneous
solution (\ref{homsolu}) proportional to $C_i$ $(i=1,2)$ and $\det
W_C=2ik^5\tau^4C_1C_2$. The homogeneous solution corresponds to
the pure gauge degree of freedom and it does not contribute to the
perturbed energy density, whereas the particular solution gives
rise to the oscillating behavior with an amplitude decaying as
$\tau^{-1}$ for small scales shown in Fig. \ref{pertnewt}.

We would like to remark that the procedure followed in order to
solve the equations for a matter dominated universe is completely
general and can be applied in any other situation where the metric
perturbations is originated by some other component.

\section{Evolution of the perturbations}

In this section we shall present the results that we obtain from a
modified version of the publicly available CAMB code
\cite{Lewis:1999bs} to compute the CMB power spectrum when
electromagnetic perturbations are taken into account. Since the
background of the electromagnetic model is the same as that of
$\Lambda$CDM, we do not need to modify the background equations of
the code, although we do have to add the evolution equation for
$A_0$. We use the first order equation (\ref{Backrho2}) so that we
can relate the background electromagnetic field directly to the
present value of its density parameter. The initial condition for
this equation is unimportant for the background evolution,
although it becomes relevant for the perturbations and it is set
by assuming a power law behavior for $A_0$. For the perturbation
equations, we add the two evolution equations for the
electromagnetic perturbations given by (\ref{perteqsyn}) and
modify the corresponding terms involving dark energy
perturbations. With these modifications, the code is ready to
compute the evolution of the perturbations in the cosmology
corresponding to the electromagnetic model and, thus, obtain the
CMB power spectrum as well as the matter power spectrum.

Before proceeding to show the obtained results, we shall discuss
what the initial conditions for the electromagnetic field
perturbations should be. A natural origin for the presence of the
new mode of the electromagnetic field on cosmological scales has
been proved to be the quantum fluctuations of such a mode during
an inflationary epoch \cite{EMdarkenergy}. In such a scenario,
only the new scalar mode can be excited because of the conformal
invariance of the usual transverse modes. The "homogeneous part"
of this scalar mode, defined as the sum of all the modes which
remain super-Hubble today, gives rise to the effective
cosmological constant whereas the modes which have already
reentered into the horizon constitute the origin of the
electromagnetic perturbations discussed in the present work. In
other words, we can split the primordial quantum fluctuations of
the scalar mode generated during inflation in a homogeneous part
comprising all the modes with $k<k_0$ (being $k_0$ the scale that
is entering into the horizon today, i.e., the present Hubble
radius) and an inhomogeneous part formed by those modes with
$k>k_0$. Notice that such a split can be performed because the
primordial power spectrum for the scalar mode is red-tilted, as
shown in \cite{EMdarkenergy}, so that the homogeneous part is
large as compared to the inhomogeneous one and this enable us to
treat the latter as a perturbation. For the mentioned scalar mode,
one can see that the longitudinal component decays with respect to
the temporal component on super-Hubble scales for a de-Sitter
inflationary epoch (in an analogous manner to that shown for the
background evolution) so that, at the end of inflation, the
amplitude of the longitudinal component would be expected to be
much smaller than the temporal one. Moreover, we have already
shown that the longitudinal components also decay with respect to
the temporal one in the radiation dominated epoch so that the
super-Hubble modes would be expected to be strongly suppressed at
the time when the initial conditions are given, which justifies to
set the initial condition for $\dA_k$ to zero. On the other hand,
the power spectrum of the quantum fluctuations generated during a
de Sitter inflation for the temporal component happens to be
scale-invariant \footnote{As we have already commented, in a more
realistic quasi-de-Sitter inflationary epoch, the power spectrum
of the temporal component becomes slightly red-tilted with a
spectral index $n_{A_0}= {\mathcal{O}}(\varepsilon)$ with
$\varepsilon$ the slow-roll parameter \cite{EMdarkenergy}.
However, we shall neglect this small spectral index in order to
set the initial conditions, being its effect expected to be
small.} so that we can set the initial condition for $\dA_{0k}$ as
$\dA_{0k}(\tau_{ini})=A\;k^{-3/2}$ with $A$ a constant depending
on the details of the inflationary epoch such as the initial
amplitude of the power spectrum after inflation or the duration of
inflation. For our purposes in this work, this constant $A$ will
play the role of a free parameter to be constrained by comparing
the CMB power spectrum produced by the model to the WMAP data. In
the same way as for the background vector field, we shall give the
initial condition for the derivative of $\dA_{0k}$ by assuming a
power-law behavior. This is justified because we know that this is
the type of evolution for the perturbations on super-Hubble
scales, where the initial conditions are given.

\begin{figure}
\begin{flushleft}
\begin{minipage}{0.45 \textwidth}
{\epsfxsize=8.7 cm\epsfbox{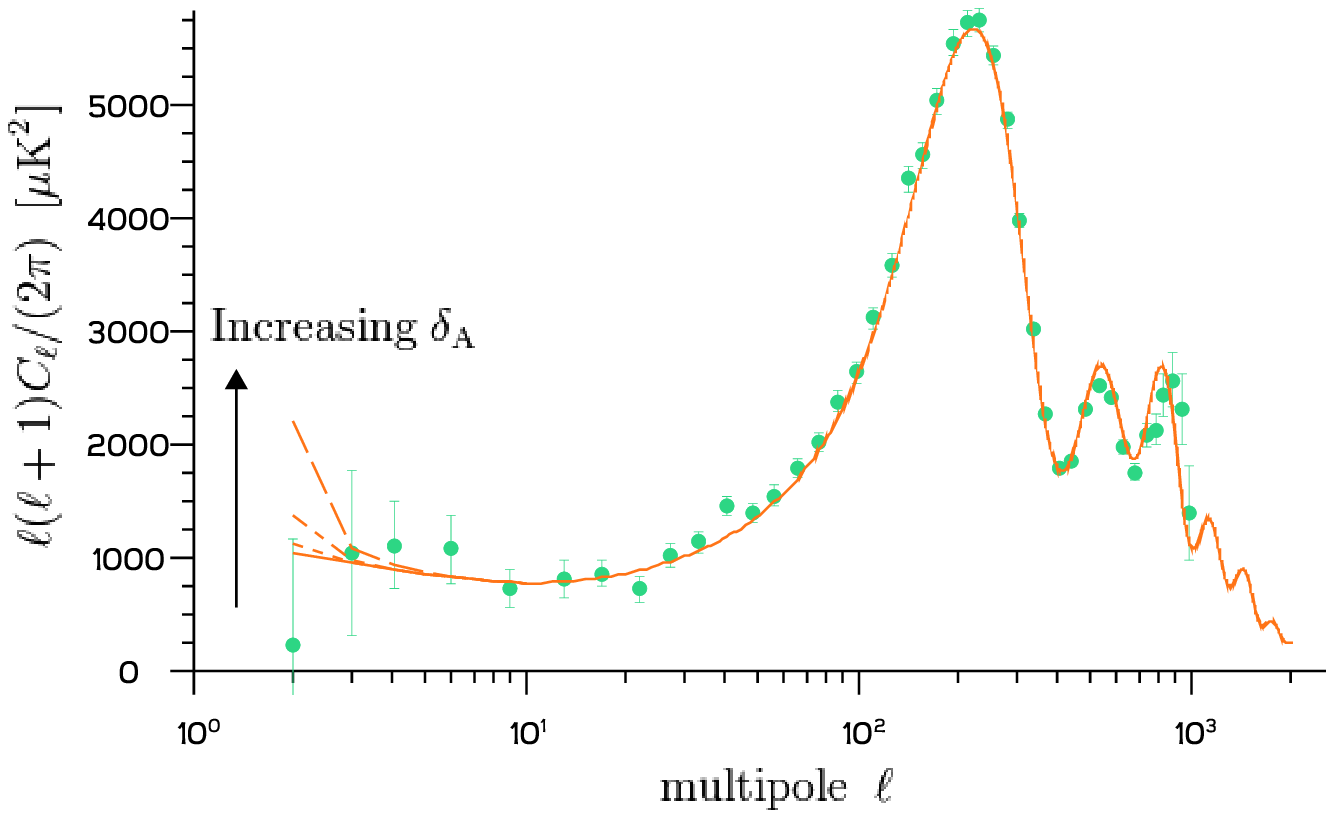}}
\end{minipage}
\hspace{0.7cm}
\begin{minipage}{0.45 \textwidth}
{\epsfxsize=8.7 cm\epsfbox{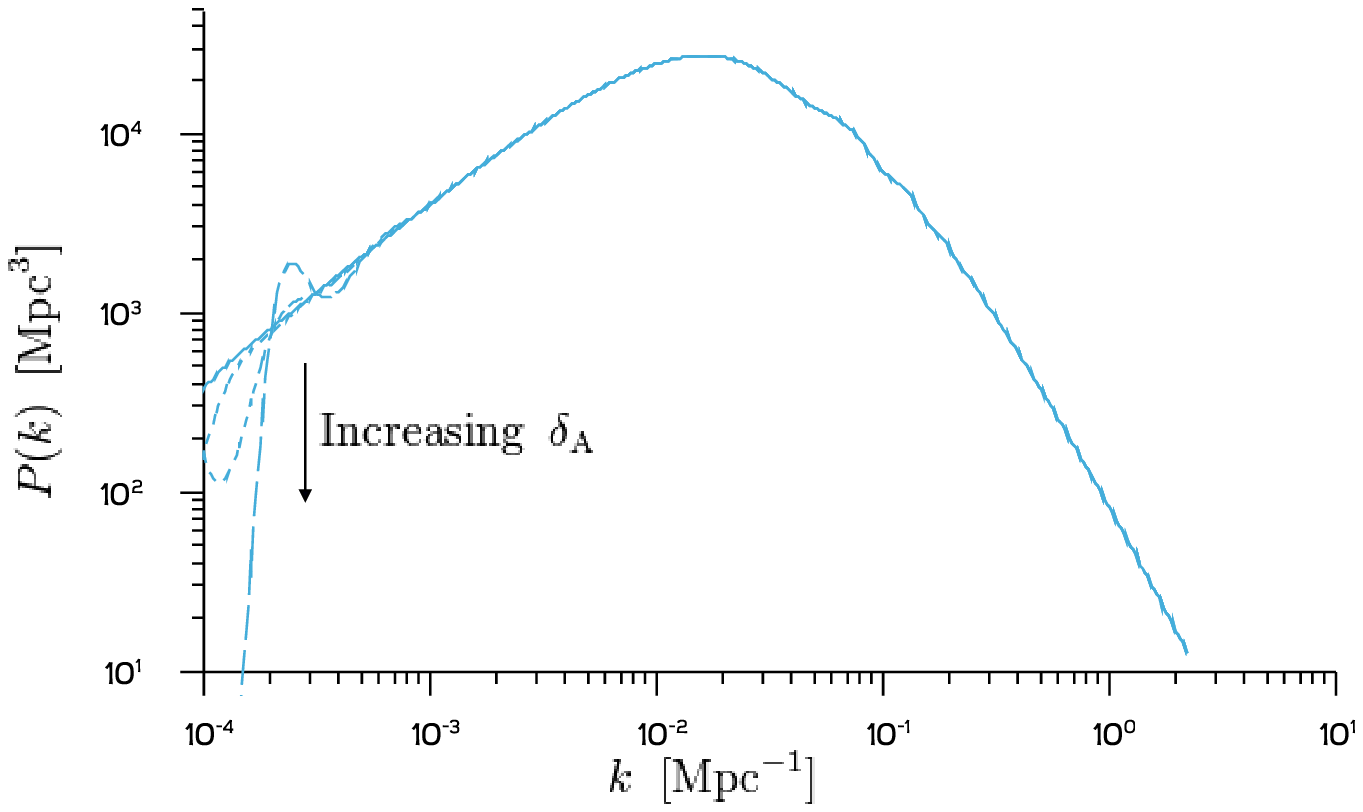}}
\end{minipage}
\end{flushleft}
\caption{In this figure we show the CMB power spectrum (left panel)
and the matter power spectrum (right panel) for both the
electromagnetic dark energy model (dotted lines) and the standard
$\Lambda$CDM model (solid lines). We have plotted several cases
with increasing values of the initial amplitude for the
electromagnetic perturbations and we see that the only
modifications appear for large scales. In particular, the CMB
quadrupole becomes excessive large, being incompatible with WMAP
data (green dots) for $\delta_A >10^{-7}$ and the small $k$
region of the matter power spectrum becomes very different from
that of $\Lambda$CDM.}
\label{spectra}
\end{figure}

\begin{figure}
\begin{center}
{\epsfxsize=17.0 cm\epsfbox{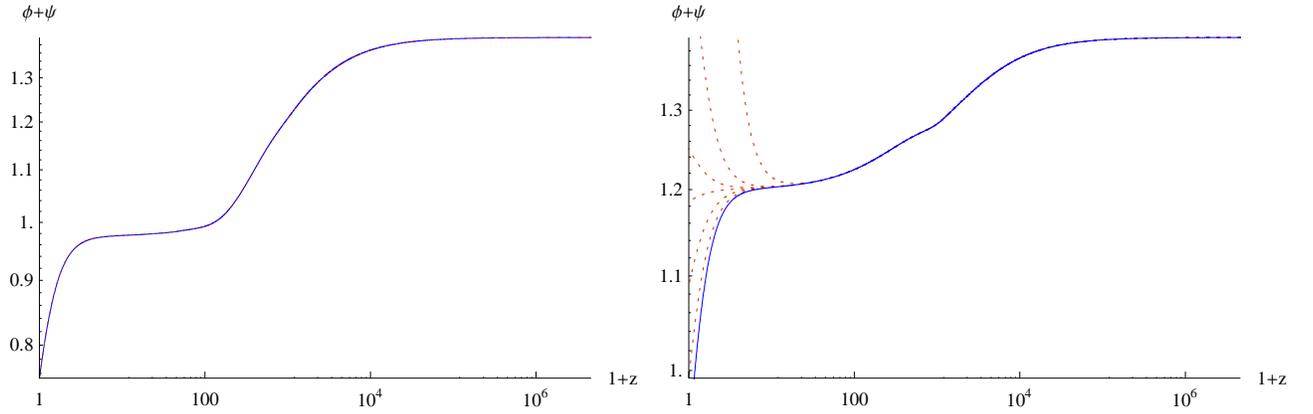}} \caption{In this figure
we show the evolution of the gravitational potentials for small
($k=5.8\times 10^{-3}$ Mpc$^{-1}$) and large ($k=1.1\times
10^{-5}$ Mpc$^{-1}$) scales in the left and right panels
respectively. We also show the evolution in a $\Lambda$CDM model
(solid blue lines) for comparison. As commented in the main text,
the evolution for the small scales is exactly the same as in the
case of a cosmological constant. In the large scales case we see
how the evolution can be very different for large enough initial
amplitudes of the electromagnetic perturbations. We plot the
electromagnetic model in dotted (red) lines and we see that the
larger the electromagnetic perturbations are, the more distinctive
the evolution of the gravitational potential
is.}\label{potentials}
\end{center}
\end{figure}

In Fig. \ref{spectra} and Fig. \ref{potentials} we show the
results obtained from the modified version of the CAMB code. The
modifications in both the CMB and matter power spectrum is
originated thanks to the fact that, unlike in the cosmological
constant case, the electromagnetic dark energy model produces
fluctuations that might modify the evolution of the gravitational
potential. In that figure, we can see that the small scales
behavior is unaffected with respect to the standard $\Lambda$CDM
case. The reason for this is that the electromagnetic
perturbations decay very rapidly once they enter into the horizon
as we have shown in the previous section so that only those
electromagnetic modes whose scales have become sub-Hubble very
recently (corresponding to the low multipoles part of the
spectrum) can contribute in a non-negligible way to the metric
perturbation evolution through Einstein equations. Moreover, since
dark energy density is negligible during decoupling, the
contribution of electromagnetic perturbations to the ordinary
Sachs-Wolfe effect is also negligible. The main effect will appear
in the late-time Integrated Sachs-Wolfe (ISW) effect as due to the
evolution of the metric perturbation. The analytical estimate of
such an effect is difficult to obtain since it requires to know
the time evolution of the metric perturbation when the
electromagnetic perturbations contribute in a non-negligible way
to the Einstein equations, so that the approximated solutions
obtained in the previous section are no longer valid. However, we
can easily understand how the ISW effect will be modified by
noticing that those modes that have become sub-Hubble very
recently can still have an appreciable amplitude and, therefore,
modify the late-time evolution of the gravitational potentials,
which gives rise to a modification of the ISW, but only for low
redshift so that the early ISW remains unaffected. Moreover, if
the corresponding mode crosses the horizon with a too large
amplitude, the modification in the gravitational potentials
evolutions might be excessively large and, thus, conflict with
observations. This is the reason why on large scales we obtain
some distinctive signatures for large enough values of the
primordial amplitude $A$. Notice that such signatures are more
apparent in the matter power spectrum. Instead of giving the
constraints in terms of the parameter $A$ we shall give the
results in terms of the more physical quantity $\delta_{A}$
defined as:
\begin{equation} \label{delta_ak}
\delta_{A}\equiv\frac{\mathcal{P}_k^{1/2}}{\rho_{\A_0}}\; .
\end{equation}
with $\mathcal{P}_k=\frac{k^3}{2\pi^2}\vert\, \delta \rho_k
\vert^2$ the power spectrum of the electromagnetic energy density
perturbations. Hence, the magnitude $\delta_A$ gives the amplitude
of the energy density fluctuations of the electromagnetic field at
a given scale $k$ relative to the homogeneous contribution. Since
$\delta \rho_k$ and $\rho_{\A_0}$ evolve in the same way on
super-Hubble scales, the quantity $\delta_{A}$ does not depend on
time as long as the corresponding mode remains super-Hubble.
Notice also that $\delta \rho_k$ contains two types of
contributions, namely: one proportional to the metric perturbation
and other proportional to the electromagnetic perturbation. Thus,
in the case when the component proportional to the electromagnetic
perturbation becomes dominant, the quantity $\delta_{A}$ becomes
scale-invariant on super-Hubble scales because
$\delta\rho_k\propto\dA_{0k}$ and $\dA_{0k}$ is proportional to
$k^{-3/2}$ due to the flatness of its primordial power spectrum.
However, if the metric perturbation contribution is dominant
$\delta_{A}$ will depend on the wave-number of the considered
mode. In any case, the upper bound that we shall obtain for
$\delta_{A}$ will show how large the primordial electromagnetic
perturbations are allowed to be in order to be compatible with CMB
measurements. In particular, we obtain that $\delta_{A}$ must
satisfy the constraint $\delta_{A}\lsim 10^{-7}$ in order not to
be in conflict with the CMB quadrupole. In fact, the overall
effect on the CMB power spectrum would be that the higher value of
$\delta_{A}$ is, the more tilted the lower multipoles part of the
spectrum becomes. In fact, since the $\delta \rho_k$ modes decay
rapidly once they reenter into the horizon (as we already
commented above) the only important effect appears for the
quadrupole and, as a consequence, the bound on $\delta_A$ is
actually a bound on such a quantity at the quadrupole scale.

On the other hand, the upper bound obtained for $\delta_{A}$ can
be linked to a variation of the Hubble parameter in a
quasi-de-Sitter inflationary epoch where the the Hubble parameter
is not exactly constant but varies slightly. To show such a link,
we have to remind that the background energy density of the
electromagnetic field is given by all the modes whose scales are
larger than the Hubble radius today, whereas the perturbations
correspond to modes whose scales are smaller than the present
Hubble radius, as commented before. In other words, the background
is given by the modes that remain super-Hubble at the present
epoch and the perturbations correspond to those modes which have
already entered into the horizon. In \cite{EMdarkenergy} we showed
that the amplitude of the electromagnetic field fluctuations (for
the temporal component) at a scale $k$ is given by $H_k$, with
$H_k$ the Hubble parameter at the time when the scale $k$ exits
from the horizon. Now, let us notice that
\begin{equation}
\delta_A\sim
\frac{H_k^2A_0\dA_{0k}}{H_I^2A_0^2}\sim\left(\frac{H_k}{H_I}\right)^3
\end{equation}
where $H_I$ is the Hubble parameter at the beginning of inflation.
Then, the constraint $\delta_A\lsim 10^{-7}$ implies that the
Hubble parameter must have reduced (at least) by a factor $\sim
200$ since the beginning of inflation until the time when the
scale of the present Hubble radius (quadrupole scale) left the
horizon, i.e., $H_{k_0}\lsim H_I/200$. Since the Hubble parameter
is proportional to the square of the inflation scale we also
obtain that the scale of inflation must reduce in a factor $\sim
15$. Notice also that this fact requires a red-tilted spectrum for
inflation, i.e., the Hubble parameter must decay throughout the
inflationary epoch. Moreover, since $\dA_{0k}\sim H_k$ (as we have
already said) we have that the fluctuations of the temporal
component must satisfy:
\begin{equation}
\frac{\dA_{0k}}{A_0}\;\lsim \;5\times10^{-3}.
\end{equation}
Then, we conclude that the electromagnetic dark energy model is
compatible with current CMB and LSS unless very large
electromagnetic initial perturbations are generated during inflation.


\section{Discussion}
\label{disc}

In the quantization of the electromagnetic field, either in the
covariant or in the path-integral formalism, one has to introduce
a gauge-fixing term in order to eliminate unphysical modes. The
form of the gauge fixing term is unique in the sense that it is
the only one that does not require the introduction of additional
scales in the action and leads to linear equations of motion,
although one may even consider nonlinear terms, in which case
Faddeev-Popov ghosts appear also in flat space-time quantum
electrodynamics. It has been proposed that the familiar, linear
gauge-fixing term might have physical consequences in curved
spacetime. One then promotes the gauge-fixing term into physical
term which is included in the fundamental theory and not
introduced during the quantization procedure as usually. In flat
space, the usual predictions of electromagnetism are retained.
However, a gravitational field may excite a scalar degree of
freedom of the electromagnetic field, which now is not a gauge
mode but has physical consequences. A remarkable result of this is
that in the homogeneous and isotropic background the temporal
component of the vector $A_\mu$ assumes an evolution that
conspires with the cosmological expansion to contribute as an
effective cosmological constant. This constant is of the order of
magnitude required by the present cosmological data given that the
field was generated during inflation occurring at the electroweak
scale.

Such a term, being not covariantly constant but having internal
dynamics, implies there will be fluctuations. We analysed these
fluctuations both analytically and numerically. At superhorizon
scales, the perturbations tend to freeze to constant value, while
after entering horizon, they decay exhibiting oscillations. As the
cosmological constant begins to contribute more significantly to
the energy density, the fluctuations of the field begin to affect
the gravitational potentials. Since this happens at recent times,
the effects to the matter power spectrum is confined to large
scales, and to the CMB at small multipoles. However, depending on
the initial amplitude of the perturbations, which in turn depends
on the details of inflation, these effects might be detectable. In
terms of the parameter (\ref{delta_ak}) we find $\delta_{A}\;
\lsim 10^{-7}$. This implies that the amplitude of fluctuations
generated for the electromagnetic field is $\lsim \;5\times 10^{-3}$.
Moreover, this fact predicts a reduction of the scale of inflation
in a factor (at least) $\sim 15$ since the beginning of
inflation until the scale of the quadrupole crossed the horizon.
Since the mode $\nabla_\mu A^\mu$ behaves as a free test scalar
field, see Eq.(\ref{box}), we expect that it acquires the spectrum
whose amplitude is given by the Hubble parameter just like that of
the inflaton field (with just a slightly different slope due to
negligible backreaction).

This shows that the model can be perfectly viable, but even more
interestingly, for some specific inflationary scenarios there
could be observable signatures from the model. Initial
fluctuations of the order $\gsim 10^{-3}$ can already be ruled
out. In addition, if a coupling existed between the inflaton and
the electromagnetic field, or if the temporal component of the
latter contributed non-negligibly to the inflationary expansion,
one would expect their spectra to be correlated, which could yet
lead to a new kind of signatures.

\vspace{0.5cm}

{\bf Acknowledgements:} This work has been  supported by
Ministerio de Ciencia e Innovaci\'on (Spain) project numbers FIS
2008-01323 and FPA 2008-00592, UCM-Santander PR34/07-15875. J.B.
wishes to thank the Institute for Theoretical Physics of the
University of Heidelberg for their hospitality.

\appendix
\section*{APPENDIX: Perturbation equations in the synchronous gauge}

In this appendix we shall derive the equations of the
perturbations and all the other expressions in the synchronous
gauge, for which the perturbed line element is:
\begin{equation}
ds^2=a(\tau)^2\left[d\tau^2-(\delta_{ij}+h_{ij})dx^idx^j\right]
\end{equation}
The scalar modes of the perturbation are usually expressed in
terms of the scalar functions $h$ and $\eta$ defined by means of
\cite{MaandB}:
\begin{eqnarray}
h_{ij}^{(s)}=\int
d^3k\left[\frac{k_ik_j}{k^2}h_k(\tau)+\left(\frac{k_ik_j}{k^2}-
\frac{1}{3}\delta_{ij}\right)6\eta_k(\tau)\right]e^{i\vec{k}\cdot\vec{x}}
\end{eqnarray}
To go from one gauge to another we perform a coordinate
transformation given by:
\begin{equation}
\hat{x}^\mu=x^\mu+\zeta^\mu
\end{equation}
As we are considering only scalar perturbations we can set
$\zeta^\mu=(\alpha,\nabla\beta)$. The relation between the scalar
metric perturbations in both gauges is as follows \cite{MaandB}:
\begin{eqnarray}
\psi_k&=&\frac{1}{2k^2}\left[h_k''+6\eta_k''+\h(h_k'+6\eta_k')\right]\\
\phi_k&=&\eta_k-\frac{1}{2k^2}\h(h_k'+6\eta_k')
\end{eqnarray}
with $\alpha_k=\beta_k'$ and
\begin{equation}
\beta_k=-\frac{1}{2k^2}(h_k+6\eta_k)
\end{equation}
The transformation for the vector field $A_\mu$ and its
energy-momentum tensor are given by $\delta_\zeta
A_\mu=-\mathcal{L}_\zeta A_\mu$ and $\delta_\zeta
T^\mu_{\;\,\nu}=-\mathcal{L}_\zeta T^\mu_{\;\,\nu}$ respectively.
From these transformation laws we obtain for the vector field
\begin{eqnarray}
\delta_\zeta \A_0=-(\alpha \A_0)'\\
\delta_\zeta \A_i=-\partial_i\alpha \A_0
\end{eqnarray}
so that the perturbations $\dA_0$ and $\dA$ in both gauges relate
as:
\begin{eqnarray}
\dA_0^{conf}&=&\dA_0^{syn}+(\alpha A_0)'-2\h\alpha A_0\nonumber\\
\A^{conf}&=&\A^{syn}+\alpha A_0
\end{eqnarray}
These relations show that the combination $\dA_0-\A+2\h A_0$ is
gauge-invariant (in the sense of cosmological perturbations).

On the other hand, for the energy-momentum tensor we obtain:
\begin{eqnarray}
\delta_\zeta T^0_{\;\,0}&=&-\alpha\, T^0_{\;\,0,0}\\
\delta_\zeta T^i_{\;\,j}&=&-\alpha\, T^i_{\;\,j,0}\\
\delta_\zeta
T^0_{\;\,i}&=&-\left(T^0_{\;\,0}-\frac{1}{3}\,T^j_{\;\,j}\right)\partial_i\alpha
\end{eqnarray}
It is very interesting to note that, given that the background
evolution of the vector field is the same as that of a
cosmological constant we have that both the unperturbed energy
density and pressure are constant and they satisfy $\rho+p=0$ so
that the energy momentum tensor remains invariant after the gauge
transformation. In fact, this is what one would expect from
Stewart-Walker lemma \cite{Stewart:1974uz}. Notice also that due to the gauge invariance
of the energy momentum tensor the results showed in Fig.
\ref{pertnewt} are also valid for the synchronous gauge.

The perturbed equations for the vector field in the synchronous
gauge are:
\begin{eqnarray}
&&\dA_{0k}''-4\h\dA_{0k}'-3k^2\dA_{0k}=-2k^2\left(2\dA_k'-5\h\dA_k\right)
-\frac{1}{2}\left[(h_k''-4\h h_k')A_0+h_k'A_0'\right]\\
&&\dA_k''-4\h\dA_k'+\left(4\h^2-2\h'-\frac{1}{3}k^2\right)\dA_k=
2\left(\frac{2}{3}\dA_{0k}'-\h\dA_{0k}\right)+\frac{1}{6}h_k'A_0\label{perteqsyn}
\end{eqnarray}
and the perturbed energy-momentum components in the synchronous
gauge are:
\begin{eqnarray}
\delta
T^0_{\;\,0}&=&\frac{1}{6a^8}\left\{\left(2\dA_{0k}'+A_0h_k'\right)A_0'+
2k^2\left[(-3\dA_{0k}+3\dA_k'-6\h\dA_k)A_0+A_0'\dA_k\right]\right\}\nonumber\\
\delta
T^i_{\;\,j}&=&\frac{1}{6a^8}\left\{\left(2\dA_{0k}'+A_0h_k'\right)A_0'+
2k^2\left[3\dA_{0k}-3\dA_k'+6\h\dA_k)A_0+A_0'\dA_k\right]\right\}\delta^i_{\;\,j}\nonumber\\
\delta
T^0_{\;\,i}&=&-\frac{ik_i}{6a^8}A_0\left[h_k'A_0+2\dA_{0k}'+2k^2\dA_k\right]
\end{eqnarray}
Condition (\ref{masspert}) remains the same in this case:
\begin{equation}
\delta(\rho_k+
p_k)=-\frac{2A_0k^2}{a^8}\left(\dA_{0k}-\dA_k'+2\h\dA_k\right)
\end{equation}
which is just a consequence of the aforementioned fact that the
combination $\dA_{0}-\dA'+2\h\dA$ does not depend on the gauge
choice.

\end{document}